# Almost Decouplability of any Directed Weighted Network Topology

Cai, Ning[1]  Khan, M. Junaid[2]

[1]College of Electrical Engineering, Northwest University for Nationalities, Lanzhou, China
[2]PN Engineering College, National University of Sciences and Technology, Karachi, Pakistan

**Abstract:** This paper introduces a conception that any weighted directed network topology is almost decouplable, which can help to transform the topology into a similar form being composed of uncoupled vertices, and thus reduce the complexity of analysis for networked dynamical systems. As an example of its application, the consensus problem of linear multi-agent systems with time-varying network topologies is addressed. As a result, a necessary and sufficient condition for uniform consensus is proposed.

## 1. Introduction

We are entering a networked world. The notion of complex and dynamical networks keeps on influencing nearly all kinds of disciplines, ranging from communication networks, power networks, biological networks, and economic networks, etc. Its impact on modern engineering and science has already been prominent, and will potentially be even far-reaching.

The graph topology of a network corresponds to certain matrix [1]. For dynamical networked systems, it is a fundamental technique to transform the network topology into similar canonical forms which are equivalent to the original network in the sense of system dynamics but are more convenient to analyze. It is usually expected that a complex network could be transformed into a similar diagonal canonical form, being composed of a number of uncoupled vertices [2]. Unfortunately, such a consideration does not always seem applicable, since a directed network is generally not guaranteed to be decouplable.

[1] Corresponding author: Cai, Ning (e-mail: caining91@tsinghua.org.cn)

This paper will show that any given weighted directed network topology is almost decouplable, despite the diagonalizability of the associated matrix.

The motivation of the current work originated from our study on consensus problems for dynamical multi-agent systems [3-6]. Consensus implicates that a system achieves certain asymptotic stability even if there are no equilibria, and it actually plays the role like equilibrium of isolated dynamical systems [3-5]. Due to its theoretical importance, the consensus problem has been paid extensive attention by scholars in the area of control theory from various perspectives; see [7-11] and references therein. Particularly, a general framework of the consensus problem for first-order systems is initially proposed in [8]. A necessary and sufficient condition for consensus of LTI first-order systems is given in [9]. Consensus problem of high-order systems is earlier addressed in [12-13]. Both observer-based dynamical protocol and robust consensus are dealt with in [14]. A noticeable approach about oblique decomposition of the state space is devised in [15-16]. The consensus condition of a class of systems with switching graph topology is shown in [17]. A method based on relative Lyapunov function is proposed in [6], which could potentially handle both consensus and non-consensus stability problems for systems composed of nonlinear or even heterogeneous agents.

It is a natural tendency to diagonalize the Laplacian matrix of the graph topology while addressing the consensus problems of multi-agent systems with linear high-order inter-agent protocols. However, due to the requirement of diagonalizability, many discussions are limited only to the case of bidirectional networks [13, 17].

We shall endeavor to reveal a fact that it is essentially not a problem whether or not a weighted directed network is decouplable. With this fact known, in many cases one can easily analyze the dynamics of multi-agent systems via the technique of decoupling the directed network through similarity transformation, with no need to consider the actual diagonalizability of associated matrix. The computational complexity might be reduced significantly, since a high-dimensional large-scale system could thus be converted into a series of independent low-dimensional subsystems. So far as we know, this idea is especially advantageous in dealing with the consensus problems of systems with uncertain or time-varying network topologies.

This paper includes two major parts. The first part is purely mathematical, and it theoretically analyzes the concept of almost decouplability. As an example to illustrate the application of the main theory, the consensus problem of linear high-order multi-agent systems with time-varying graph topologies is concerned in the second part of the paper.

It is worthy to mention that some of the propositions may seem lesser novel alone. Some analogous arguments could also exist in the literature, e.g. [18]. However, by systematically clarifying and integrating the relevant considerations, both from Mathematics and systems theory viewpoints, a novel and effective methodology can be yielded to deal with many problems in dynamical networked systems, especially stability problems. This is the essential contribution of the current paper.

The rest of this paper is organized as follows. Section 2 depicts the almost decouplability concept, providing the main theoretical result. Section 3 deals with the stability problem of general dynamical systems as preparation for the next section. Section 4 discusses the uniform consensus for multi-agent systems, based on the results shown in previous sections. Finally, the concluding remarks form Section 5.

## 2. Almost decouplatility of directed weighted networks

As the most important part of the paper, the current section expounds the almost decouplability conception. We can see that the meaning of this conception is twofold, which will be reflected by Proposition 1 & Theorem 1, respectively: 1) Without considering the limitation of structure, the Lebesgue measure for the set of undiagonalizable square matrices is zero; 2) For any given directed weighted graph, there exists an arbitrarily close graph with diagonalizable Laplacian matrix, which has a prescribed structure limitation.

*Proposition 1:* Any asymmetric real-valued square matrix $L \in R^{N\times N}$ is almost diagonalizable, i.e. the Lebesgue measure for the set of undiagonalizable matrices in $R^{N\times N}$ is zero.

The subsequent theoretical grounds are essential for proving Proposition 1, which are quoted from linear algebra.

*Definition 1* (Resultant) [19]: Suppose that $f(x) = a_0x^m + a_1x^{m-1} + \cdots + a_m$ ( $a_0 \neq 0$ ) and $g(x) = b_0x^n + b_1x^{n-1} + \cdots + b_n$ ( $b_0 \neq 0$ ), with their roots being $\alpha_1, \alpha_2, \cdots, \alpha_m$ and $\beta_1, \beta_2, \cdots, \beta_n$ respectively. The resultant of the two polynomials is defined as follows:

$$R(f,g) = a_0^n b_0^m \prod_{i=1}^{m} \prod_{j=1}^{n} (\alpha_i - \beta_j)$$

With the following formula [19], one can directly compute the resultant of two polynomials by their coefficients:

$$R(f,g) = \Delta(m,n) = \begin{vmatrix} a_0 & a_1 & \cdots & a_m & & & & \\ & a_0 & a_1 & \cdots & a_m & & & \\ & & & & \ddots & & & \\ & & & & a_0 & a_1 & \cdots & a_m \\ b_0 & b_1 & \cdots & b_n & & & & \\ & b_0 & b_1 & \cdots & b_n & & & \\ & & & & \ddots & & & \\ & & & & b_0 & b_1 & \cdots & b_n \end{vmatrix}$$

*Lemma 1* [19]: An irreducible polynomial $p(x)$ is a multiple factor of polynomial $f(x)$ if and only if $p(x)$ is a common factor of $f(x)$ and $f'(x)$, where $f'(x)$ is the derivative of $f(x)$.

*Lemma 2* [19]: There exists a common factor of polynomials $f(x)$ and $g(x)$ if and only if the resultant $R(f,g)=0$.

*Proof of Proposition 1:* Consider any arbitrary matrix $L = \left[ l_{ij} \right] \in R^{N \times N}$. Suppose that its characteristic polynomial is

$$f(x) = |xI - L| = x^N + a_1 x^{N-1} + \cdots + a_N \quad (1)$$

with the coefficient vector $\begin{bmatrix} a_1 & a_2 & \cdots & a_N \end{bmatrix}^T \in R^N$. The derivative of $f(x)$ is

$$f'(x) = N x^{N-1} + a_1 (N-1) x^{N-2} + \cdots + a_{N-1}$$

According to Lemmas 1 & 2, there exists a multiple factor of $f(x)$ if and only if $R(f, f') = 0$, i.e. the coefficients of $f(x)$ must satisfy the following equation:

$$\begin{vmatrix} 1 & a_1 & \cdots & a_N & & & & \\ & 1 & a_1 & \cdots & a_N & & & \\ & & & & \ddots & & & \\ & & & & 1 & a_1 & \cdots & a_N \\ N & a_1(N-1) & \cdots & a_{N-1} & & & & \\ & N & a_1(N-1) & \cdots & a_{N-1} & & & \\ & & & & \ddots & & & \\ & & & & N & a_1(N-1) & \cdots & a_{N-1} \end{vmatrix} = 0 \quad (2)$$

Evidently, the set of variables $a_1, \cdots, a_N$ satisfying the above algebraic equation forms a manifold of dimension $N-1$.

Equation (2) determines the relationship between the coefficient vector $\begin{bmatrix} a_1 & a_2 & \cdots & a_N \end{bmatrix}^T$ and the values of entries in $L$, which are $N$ scalar functions:

$$\begin{cases} a_1 = \varphi_1(l_{11}, l_{12}, \cdots, l_{NN}) \\ a_2 = \varphi_2(l_{11}, l_{12}, \cdots, l_{NN}) \\ \quad \vdots \\ a_N = \varphi_N(l_{11}, l_{12}, \cdots, l_{NN}) \end{cases} \tag{3}$$

Substituting $a_1, \cdots, a_N$ in (2) by (3) will result in an equation taking the following form:

$$\Gamma(l_{11}, l_{12}, \cdots, l_{NN}) = 0 \tag{4}$$

with $\Gamma(\bullet)$ being a scalar function. It is evident that the set of matrices that satisfy (4) forms a manifold of lesser dimension in $R^{N\times N}$ and therefore its Lebesgue measure is zero. All the matrices in $R^{N\times N}$ except this set have distinct eigenvalues and are surely diagonalizable. □

*Example 1:* Consider any matrix in $R^{2\times 2}$, which is

$$\begin{bmatrix} a & b \\ c & d \end{bmatrix}$$

with $a, b, c$ and $d$ as some arbitrary real numbers. Its characteristic polynomial is

$$\begin{vmatrix} x-a & -b \\ -c & x-d \end{vmatrix} = x^2 - (a+d)x + ad - bc$$

The polynomial has multiple roots if and only if

$$(a+d)^2 = 4(ad-bc)$$

which determines a manifold of measure 0 in $R^{2\times 2}$.□

*Remark 1:* The zero Lebesgue measure for the set of undiagonalizable matrices also implies that the diagonalizable set is dense. However, the former condition is stronger than the mere density. As a well-known instance, the set of rational numbers is dense in $R$ whereas its measure is still zero.

*Remark 2:* The analysis in practical scenarios can be much convenient, with the knowledge that the Lebesgue measure of the set of undiagonalizable matrices is zero. Because deviations always exist between models and the real systems, usually a matrix encountered in an engineering scenario, such as an adjacency matrix or Laplacian matrix of a specific network topology, can be regarded as being diagonalizable according to Proposition 1. In most cases, it is not necessary to consider the undiagonalizable exceptions.

If the graph topology of a network is denoted by its adjacency matrix for the sake of analysis, then there is a one-to-one correspondence between any graph and a matrix in real number domain; thus, Proposition 1 is sufficient to interpret the idea that the graph topology is almost decouplability. However, while concerning the dynamical multi-agent systems with linear inter-agent protocols, Laplacian matrices are more frequently adopted to represent the graph topologies. Laplacian matrix is a specific type of matrix with structure limitation, i.e. the sum of elements in any row is zero. It cannot guarantee that Proposition 1 is still valid if the matrix has certain structure limitation. The next theorem deals with such a situation. The lemmas that follow are theoretical preparations for the theorem.

*Definition 2* (Laplacian Matrix) [1]: The Laplacian matrix of a given directed weighted network topology is

$$L = D - W$$

where $W$ is the adjacency matrix and $D$ is the in-degree matrix with $D = diag(W\phi)$. Note that $\phi = \begin{bmatrix} 1 & 1 & \cdots & 1 \end{bmatrix}^T$ denotes a vector with all entries being 1.

*Lemma 3:* $\phi$ is an eigenvector of any Laplacian matrix $L$ with the associated eigenvalue being 0.

*Proof:* Because $D\phi = diag(W\phi)\phi = W\phi$, $L\phi = 0$.□

Lemma 3 can be regarded as a structural limitation or a criterion to check whether a matrix is a Laplacian matrix of certain network topology.

*Lemma 4* [20]: (Schur's Unitary Triangularization Theorem) With given $A \in R^{n\times n}$, there exists a unitary matrix $U$ such that $UAU^*$ is upper triangular.

*Lemma 5* [20]: If two matrices $A = [a_{ij}]$ and $B = [b_{ij}]$ in $R^{n\times n}$ are unitarily equivalent, i.e. there is a unitary matrix $U$ such that $UAU^* = B$, then

$$\sum_{i,j=1}^{n} a_{ij}^2 = \sum_{i,j=1}^{n} b_{ij}^2$$

*Theorem 1:* For any directed graph $G$ of $N$ th order and any value $\varepsilon > 0$, there exists a graph $G(\varepsilon)$ that is decouplable, i.e. its Laplacian matrix $L(G(\varepsilon)) = [l_{ij}(\varepsilon)]$ is diagonalizable, meanwhile

$$\sum_{i,j=1}^{N}(l_{ij}-l_{ij}(\varepsilon))^2<\varepsilon$$

*Proof:* Suppose that the eigenvalues of $L$ are $0,\lambda_2,...,\lambda_N$. Let $U\in R^{N\times N}$ be unitary such that $U^*LU$ is upper triangular. Select an upper triangular matrix $E\in R^{N\times N}$ in which the elements satisfy

$$\left|e_{ij}\right|<\sqrt{2\varepsilon/N(N+1)}\quad(i=1,2,...,N\;\;j=i,...,N)\tag{5}$$

such that $\lambda_2+e_{22},...,\lambda_N+e_{NN}$ are different from each other. Besides, let the elements in $E$ also jointly satisfy (6):

$$UEU^*\phi=0\tag{6}$$

Equation (6) is a system of equations with $N$ equations and $(1+N)N/2$ independent variables, so there must exist feasible solutions since $N\le N(N+1)/2$. With such an $E$ satisfying (6), let $L(G(\varepsilon))=L+UEU^*$, then $L(G(\varepsilon))$-$L$=$UEU^*$ and according to Lemma 5 and (5),

$$\sum_{i,j=1}^{N}(l_{ij}-l_{ij}(\varepsilon))^2=\sum_{i=1}^{N}\sum_{j=i}^{N}e_{ij}^2<\varepsilon$$

Evidently, $L(\varepsilon)$ is diagonalizable because its spectrum is $\{0,\lambda_2+e_{22},...,\lambda_N+e_{NN}\}$, without identical eigenvalues. □

*Remark 3:* Proposition 1 & Theorem 1 play different roles, and are compensating each other. They jointly constitute the interpretation for the almost decouplability conception proposed in this paper. Proposition 1 implicates a fact that the set of diagonalizable matrices is dense, including most of the matrices; while Theorem 1 concentrates on the diagonalizability of Laplacian matrices, which is the case with structure limitation.

The idea of almost decouplability elaborated in this section may seem obvious, but it can truly help to reduce the complexity of analysis. The next two sections will illustrate the effectiveness of this idea by an example on the consensus problem of dynamical multi-agent systems with time-varying network topologies.

# 3. Uniform asymptotic stability of slightly perturbed time-varying dynamical systems

This section discusses some general problems in control theory, leading to two main propositions which collectively imply a fact that the uniform asymptotic stability is robust to sufficiently minor uncertainties of system dynamics across the entire time horizon. As a result, it is feasible to study the stability of a dynamical system via replacing it by another system with just slight variation, but with remarkably simplified analysis. The current section is essentially independent of Section 2, and its main purpose is to provide necessary theoretical basis to support the analysis on the consensus problem in the next section.

*Definition 3* (Positive Definite Function): A real-valued, continuously differentiable function $f(x)$ is said to be *positive definite* in a neighborhood of the origin $D$ if $f(0)=0$ and $f(x)>0$ for any nonzero $x \in D$. A function $f(t,x)$ ($[0,\infty)\times D \to R$) is said to be *positive definite* if $f(t,0)=0$ and $f(t,x)\geq v(x)$ in $[0,\infty)\times D$ for some positive definite function $v(x)$.

*Lemma 6* [21]: Let $x=0$ be a uniformly asymptotically stable equilibrium point for the nonlinear system

$$\dot{x} = f(t,x)$$

where $f:[0,\infty)\times D \to R^n$ ($D=\left\{x\in R^n \,\big|\, \|x\|<r\right\}$) is continuously differentiable both in $t$ and $x$. Then, there exist a positive real value $r_0$ and a continuously differentiable function $V:[0,\infty)\times D_0 \to R^n$ ($D_0=\left\{x\in R^n \,\big|\, \|x\|<r_0\right\}$) that satisfies the following series of inequalities

$$\alpha_1(\|x\|) \leq V(t,x) \leq \alpha_2(\|x\|) \tag{7}$$

$$\frac{\partial V}{\partial t} + \frac{\partial V}{\partial x} f(t,x) \leq -\alpha_3(\|x\|) \tag{8}$$

$$\left\|\frac{\partial V}{\partial x}\right\| \leq \alpha_4(\|x\|) \tag{9}$$

where $\alpha_1$, $\alpha_2$, $\alpha_3$, and $\alpha_4$ are class $K$ functions.

*Proposition 2:* Suppose that $x=0$ is a uniformly asymptotically stable equilibrium point of a nonlinear time-varying system $\dot{x} = f(t,x)$, where $f(t,x)$ is continuously differentiable both in $t$ and $x$, then there must exist a positive definite function $\gamma(t,x)$ such that $x=0$ is also uniformly asymptotically stable for all the systems:

$$\dot{x} = f(t,x) + \delta(t,x) \tag{10}$$

where $\delta(t,x)$ is any vector function in $R^n$ that satisfies

$$\|\delta(t,x)\| \le \gamma(t,x)$$

*Proof:* It is simple to verify that the origin $x=0$ is also an equilibrium point for any system (10). According to Lemma 6, there exists a continuously differentiable function $V(t,x)$ satisfying inequalities (7)~(9). Let $V(t,x)$ be the Lyapunov function candidate for system (10). Then its derivative along the trajectory of (10) is

$$\begin{aligned}\dot{V} &= \frac{\partial V}{\partial t} + \frac{\partial V}{\partial x}\dot{x} \\ &= \frac{\partial V}{\partial t} + \frac{\partial V}{\partial x} f(t,x) + \frac{\partial V}{\partial x}\delta(t,x)\end{aligned} \tag{11}$$

In (11) above,

$$\left|\frac{\partial V}{\partial x}\delta(t,x)\right| \le \left\|\frac{\partial V}{\partial x}\right\| \|\delta(t,x)\|$$

Let $\gamma(t,x)$ be a positive definite function satisfying

$$\gamma(t,x) < \alpha_3(\|x\|) / \left\|\frac{\partial V}{\partial x}\right\| \quad (x \ne 0) \tag{12}$$

Then, because $\|\delta(t,x)\| \le \gamma(t,x)$,

$$-\alpha_3(\|x\|) < \frac{\partial V}{\partial x}\delta(t,x) < \alpha_3(\|x\|) \tag{13}$$

Considering both (13) and (8), it is evident to know that (11) is negative for $\forall x \ne 0$. Thus, $x=0$ is uniformly asymptotically stable for any system (10). □

*Proposition 3:* Suppose that $x=0$ is an equilibrium point of a nonlinear time-varying system $\dot{x} = f(t,x)$, then there must exist a positive definite function $\gamma(t,x)$ such that $x=0$ is uniformly asymptotically stable for system $\dot{x} = f(t,x)$ if and only if $x=0$ is also uniformly asymptotically stable for system

$$\dot{x} = f(t,x) + \delta(t,x)$$

where $\delta(t,x)$ is any vector function with proper dimension satisfying

$$\|\delta(t,x)\| \le \gamma(t,x)$$

*Proof:* According to Proposition 2, for any nonlinear system with origin its uniformly asymptotically stable equilibrium point, there is a positive definite function indicating the bound of permitted variation preserving stability. Thus, there is a map between a system and a positive definite function:

$$\dot{x} = f(t,x) \to \gamma(t,x) \tag{14}$$

Assume that $x=0$ is uniformly asymptotically stable for a specific system $\dot{x} = f_0(t,x)$, then there exists a positive definite function $\gamma_0(t,x)$ such that $x=0$ is uniformly asymptotically stable for the set of systems

$$\left\{\dot{x} = f_0(t,x) + \delta(t,x) \,\middle|\, \|\delta(t,x)\| < \gamma_0(t,x)\right\} \tag{15}$$

Each element in the above set can also induce a positive definite function, e.g.

$$\dot{x} = f_0(t,x) + \delta_0(t,x) \to \hat{\gamma}_0(t,x) \tag{16}$$

with $\delta_0(t,x)$ being a specific vector function satisfying $\|\delta_0(t,x)\| < \gamma_0(t,x)$. As a result, $x=0$ is uniformly asymptotically stable for the set of systems

$$\left\{\dot{x} = f_0(t,x) + \delta_0(t,x) + \Delta(t,x) \,\middle|\, \|\Delta(t,x)\| < \hat{\gamma}_0(t,x)\right\}$$

if $x=0$ is uniformly asymptotically stable for the system $\dot{x} = f_0(t,x) + \delta_0(t,x)$. It is simple to construct a positive definite function $\gamma(t,x)$ satisfying

$$\gamma(t,x) \le \hat{\gamma}_0(t,x)$$

for all such $\hat{\gamma}_0(t,x)$ induced by map (16) from elements in the set (15). Evidently, $x=0$ is uniformly asymptotically stable for the specific system $\dot{x} = f_0(t,x)$ if and only if $x=0$ is uniformly asymptotically stable for all the elements in

$$\left\{\dot{x} = f_0(t,x) + \delta(t,x) \,\middle|\, \|\delta(t,x)\| < \gamma(t,x)\right\}$$ □

Combining the theoretical results in Sections 2 & 3, the methodology for stability analysis of networked systems based on almost decouplability is clearly established here. The process can be summarized as follows. First, let an approximate network topology $G(\varepsilon)$ with diagonalizable Laplacian matrix or adjacency matrix replace the original $G$, no matter whether it is actually diagonalizable. Then, decouple the network via a nonsingular transformation and a system composed of independent low-dimensional subsystems is thus derived, which has the same stability with the original system.

# 4. Uniform consensus of multi-agent systems with time-varying network topologies

The major purpose of this section is to provide an example to show the effectiveness of the main theory introduced previously. As the example, the consensus problem will be analyzed for dynamical multi-agent systems with time-varying network topologies, by approaches based on the idea of almost decouplability. A necessary and sufficient condition will be derived for the uniform consensus.

The dynamic multi-agent system model concerned here comprises of $N$ agents, and each agent has dynamics of $d$th order. The state of agent $i$ is denoted by $x_i = [x_{i1}, x_{i2}, ..., x_{id}]^T \in R^d$ . The communication architecture among agents is represented by a graph topology $G$ of order $N$, with each agent corresponding to a vertex. The arc weight of $G$ between agents $i$ and $j$ is denoted by $w_{ij} \geq 0$, which can be regarded as the strength of information link. The graph $G$ in this section is time-varying, which is sufficiently smooth in $t$ and can be denoted by its adjacency matrix $W(t)$:

$$G : W = \begin{bmatrix} w_{11} & w_{12} & \cdots & w_{1N} \\ w_{21} & w_{22} & \cdots & w_{2N} \\ \vdots & \vdots & & \vdots \\ w_{N1} & w_{N2} & \cdots & w_{NN} \end{bmatrix}$$

Suppose the dynamics of each agent can be described as:

$$\dot{x}_i = Ax_i + F\sum_{j=1}^{N} w_{ij}(t)(x_j - x_i) \quad (i \in \{1, 2, ..., N\}) \tag{17}$$

where $A \in R^{d\times d}$ and $F \in R^{d\times d}$ . If a stack of state vectors is defined as $x = \begin{bmatrix} x_1^T & x_2^T & \cdots & x_N^T \end{bmatrix}^T$, then the system dynamics can be described by

$$\dot{x} = (I_N \otimes A - L(t) \otimes F)x \tag{18}$$

with $L(t)$ being the time-varying Laplacian matrix of graph $G$ .

As the theme of the current section, consensus is essentially a stability problem. The following definitions clearly indicate the relationship between consensus and stability. With these definitions, the two equivalent terms *consensus* and *asymptotic swarm stability* will be used indiscriminately later.

*Definition 4* [4]: (Swarm Stability) For a time-varying dynamical multi-agent system with $x_1, ..., x_N \in R^d$ as the states of $N$ agents, if for $\forall \varepsilon > 0$ , $\exists \delta(\varepsilon) > 0$ , *s.t.* $\|x_i(t) - x_j(t)\| < \varepsilon$ ( $t > 0$ ) as $\|x_i(0) - x_j(0)\| < \delta(\varepsilon)$ ( $\forall i, j \in \{1, 2, ..., N\}$ ), then the system is uniformly *swarm stable*. If $\lim_{\varepsilon\to\infty} \delta(\varepsilon) = \infty$ , the system is globally uniformly swarm stable.

*Definition 5* [4]: (Asymptotic Swarm Stability) If a time-varying dynamical multi-agent system is globally uniformly swarm stable and for $\forall \varepsilon, c > 0$ , $\exists T = T(\varepsilon, c) > 0$ *s.t.*

$$\|x_i(t) - x_j(t)\| < \varepsilon$$

as $t > T(\varepsilon, c)$ and $\|x_i(0) - x_j(0)\| < c$ ( $\forall i, j \in \{1, 2, ..., N\}$ ), then the system is globally uniformly *asymptotically swarm stable*.

*Lemma 7* [9]: The Laplacian matrix $L$ of a directed graph $G$ has exactly a single zero eigenvalue $\lambda_1 = 0$ iff $G$ has a spanning tree, with the corresponding eigenvector $\phi = [1 \ \ 1 \ \ \cdots \ \ 1]^T$.

*Lemma 8* [4]: For the time-varying dynamic multi-agent system (17), if it is uniformly asymptotic swarm stable, then as $t \to \infty$, the trajectory of each agent tends to be regulated by the equation $\dot{\xi} = A\xi$.

*Lemma 9* [4]: For system (17), if $A$ is Hurwitz, then the system is uniformly asymptotically swarm stable iff it is uniformly asymptotically Lyapunov stable. Besides, the consensus state must be zero.

*Theorem 2:* For the multi-agent system (17), with $\lambda_1 = 0, \lambda_2(t), \cdots, \lambda_N(t) \in C$ the time-varying eigenvalues of $L(G(t))$, if $A$ is unstable, the overall system is uniformly asymptotically swarm stable iff both 1) and 2) below are true:

1) $\exists T > 0$ *s.t.* there do not exist any two values $t_1$ and $t_2$ ($t_2 > t_1 > T$) while $G(t)$ includes no spanning tree during $t \in [t_1, t_2]$;
2) All the following $d$-dimensional dynamic systems
$$\dot{\xi} = (A - \lambda_i(t)F)\xi \quad (\lambda_i \neq 0)$$
are uniformly asymptotically stable.

If $A$ is Hurwitz, then the multi-agent system is uniformly asymptotically swarm stable iff 2) is true.

*Proof:* Part I ( $A$ is unstable)

Assume that 1) does not hold but the multi-agent system is still uniformly asymptotically swarm stable. Suppose that $G$ has no spanning tree for $t \in [t_1, t_2]$, where $t_1$ and $t_2$ can be arbitrarily large, then $G$ must contain $k \geq 2$ distinct subgraphs $\hat{G}_1$, $\hat{G}_2$ ,…, $\hat{G}_k$ , each receiving no information during $t \in [t_1, t_2]$. According to Lemma 8, as $t$ is sufficiently large, the consentaneous trajectory $\xi_1(t)$ of the agents associated with $\hat{G}_1$ approaches the solution of equation $\dot{\xi}_1 = A\xi_1$, whereas the consentaneous trajectory $\xi_2(t)$ associated with $\hat{G}_2$ approaches the solution of $\dot{\xi}_2 = A\xi_2$. The two trajectories are independent because there is no information exchanged. Thus, $\dot{\xi}_1 - \dot{\xi}_2 = A(\xi_1 - \xi_2)$. Since $A$ is unstable, $\|\xi_1 - \xi_2\|$ tends to increase during $t \in [t_1, t_2]$. This contradicts the assumption that the multi-agent system is asymptotically swarm stable. Therefore, 1) is a necessary condition. Note that the condition $G(t)$ possessing no spanning tree is permitted at any particular instant $t>T$ due to the continuity of the system.

Suppose that $\lambda_1=0,\lambda_2(t)...,\lambda_N(t)\in C$ are the eigenvalues of $L(t)$. According to Proposition 1 and Theorem 1, let an approximate $G(\varepsilon,t)$ ($\varepsilon\to 0$) replace $G(t)$, with $L(\varepsilon,t)$ replacing $L(t)$, which is uniformly diagonalizable. Hence, an alternative system of (18) is derived as:

$$\dot{x}=(I_N\otimes A-L(\varepsilon,t)\otimes F)x \tag{19}$$

Considering the continuity of the eigenvalues of matrices, one can infer that the eigenvalues of $L(\varepsilon,t)$ also approach $\lambda_1=0,\lambda_2(t)...,\lambda_N(t)$ as $\varepsilon\to 0$.

Now let us consider the swarm stability of (19). Suppose that $T(t)L(\varepsilon,t)T^{-1}(t)=\Lambda(t)$ and let $\tilde{x}=(T(t)\otimes I_d)x$, then (19) is transformed into

$$\dot{\tilde{x}}=(I_N\otimes A-\Lambda(t)\otimes F)\tilde{x} \tag{20}$$

According to Propositions 2 & 3, the uniform asymptotic stability of system (20) is equivalent to that of system (18), with $\Lambda(t)=diag(0,\lambda_2(t),\lambda_3(t),\cdots,\lambda_N(t))$. The structure of $I_N\otimes A-\Lambda(t)\otimes F$ is of the form

$$I_N\otimes A-\Lambda(t)\otimes F=\begin{bmatrix} A & 0 & \cdots & 0 \\ 0 & A-\lambda_2(t)F & & \vdots \\ \vdots & & \ddots & 0 \\ 0 & \cdots & 0 & A-\lambda_N(t)F \end{bmatrix} \tag{21}$$

Let the auxiliary variables

$$\eta_i=\sum_{j=1}^{N} w_{ij}(\varepsilon,t)(x_j-x_i) \quad (i=1,2,...,N)$$

and their stack vector be $\eta=\begin{bmatrix}\eta_1^T & \eta_2^T & \cdots & \eta_N^T\end{bmatrix}^T$. It follows that

$$\eta=(L(\varepsilon,t)\otimes I_d)x \tag{22}$$

With condition 1), according to Lemma 7, any vector in the null space of $L(\varepsilon,t)\otimes I_d$ is of the form $\phi\otimes\xi$ when $t>T$, where $\phi=\begin{bmatrix}1 & 1 & \cdots & 1\end{bmatrix}^T\in R^N$ and $\xi\in R^d$. Therefore, $x_1=x_2=\cdots=x_N$ in (19) iff $\eta=0$ when $t>T$. Because of (21) and $x=(T^{-1}\otimes I_d)\tilde{x}$,

$$(T\otimes I_d)\eta=(TL(\varepsilon)\otimes I_d)x=(TL(\varepsilon)\otimes I_d)(T^{-1}\otimes I_d)\tilde{x}=(\Lambda\otimes I_d)\tilde{x} \tag{23}$$

According to Lemma 7, all the values $\lambda_2,\cdots,\lambda_N$ are nonzero as $t>T$, thus from the linear equation (23), one can infer that $\eta\to 0$ iff $\tilde{x}_2\to\cdots\to\tilde{x}_N\to 0$ with considering the structure of $\Lambda(t)$. By (21), it is clearly known that $\tilde{x}_2,\tilde{x}_3,\cdots,\tilde{x}_N\to 0$ iff condition 2) holds. According to Propositions 2 & 3, the uniform asymptotic stability of (19) is identical with that of (18); therefore condition 2) is also appropriate for the system (18).

Part II ( $A$ is Hurwitz)

As $A$ is Hurwitz, according to Lemma 9, the system is uniformly asymptotically swarm stable iff it is uniformly asymptotically Lyapunov stable. The stability of system (17) is equivalent to that of (21). It is clear that (21) is uniformly asymptotically stable iff condition 2) is true.□

*Remark 4:* If without the almost decouplability conception introduced in previous sections, it would be difficult or even impossible to handle the proof and obtain the result of Theorem 2, since the network topology here is not only asymmetric but also time-varying.

*Remark 5:* Admittedly, Theorem 2 is relatively theoretic, since eigenvalues are usually not easy to obtain. Its main role is to demonstrate the effectiveness of the approach based on the idea of almost decouplability. Nonetheless, this theorem itself still has theoretical significance. First, it provides a necessary and sufficient condition without additional technical hypothesis. Second, the consensus of a large-scale system is transformed into the stability of a series of independent low-dimensional systems, and the computational complexity is thus reduced significantly. Besides, this result separates the consideration of network topology from the dynamics of agents.

# 5. Conclusions

This paper introduces a conception that any weighted directed network topology is almost decouplable, i.e. there always exists an arbitrarily close digraph which can be transformed into a similar form, consisting of independent vertices. This conception could help to reduce the complexity of analysis because a large scale networked system can thereby be decoupled into numerous independent low-dimensional systems through similarity transformations. As an example to illustrate the validity of the methodology proposed, the uniform consensus problem of dynamical multi-agent systems with time-varying network topologies is discussed, leading to establishing a necessary and sufficient condition. Evidently, the almost decouplability conception should also be applicable in other problems, e.g. the robust consensus problem, which will be investigated in our future work.

# Acknowledgments

This work is supported by Program for Young Talents of State Ethnic Affairs Commission (SEAC) China (Grant [2013] 231), and by National Natural Science Foundation (NNSF) of China (Grants 61174067, 61263002, & 61374054).